\documentstyle[12pt,epsfig]{article}
\begin{document}
\noindent
\begin{center}
{\Large {\bf Interacting Early Dark Energy  }}\\
\vspace{2cm}
 ${\bf Yousef~Bisabr}$\footnote{e-mail:~y-bisabr@sru.ac.ir.}\\
\vspace{.5cm} {\small{Department of Physics, Shahid Rajaee Teacher
Training University,
Lavizan, Tehran 16788, Iran}}\\
\end{center}
\begin{abstract}
We explore a model of interacting Early Dark Energy (EDE) in which a minimally coupled scalar field, representing the EDE, interacts with the radiation sector through an exponential coupling function in the radiation-dominated era. This framework can be inspired by modified theories of gravity, including $f(R)$ gravity, the Einstein frame representation of Brans-Dicke theory, and chameleon gravity. Our findings reveal that the traditional law of radiation conservation is altered to $\rho_{\gamma}\propto a^{-4+\epsilon}$, where the parameter $\epsilon$ measures the rate of energy transfer between radiation and EDE. Assuming a constant energy transfer, we show that the scalar field behaves as $\phi\propto\ln a$, indicating that $\phi$ diverges as $a$ approaches zero. Additionally, we demonstrate that the interacting scalar-photon system behaves similar to an effective cosmological constant in the early stages of evolution of the Universe.
Moreover, by solving the conservation equation associated with the scalar field, we derive an analytical expression for the ratio $r=\rho_{\phi}/\rho_{\gamma}$. Our results indicate that $r$ diminishes as the Universe expands, which is essential for a successful EDE model. Our investigation into the parameter space confirms that the expected behavior of $r$ during recombination aligns with contemporary cosmological data. These insights underscore crucial aspects necessary for any feasible EDE model and present exciting possibilities for resolving the Hubble tension.

\end{abstract}

~~~~~~~~~~~~~~~~~~~~~~~~~~~~~~~~~~~~~~~~~~~~~~~~~~~~~~~~~~~~~~~~~~~~~~~~~~~~~~~~~~~~~~~~~~~~~~~~~~~~~
\section{Introduction}
The $\Lambda$CDM model stands as the predominant paradigm in contemporary cosmology, providing a robust framework for understanding the large-scale structure and development of the Universe. This model is built on two key elements: Firstly, the cosmological constant $\Lambda$,   which symbolizes dark energy and propels the accelerated expansion of the Universe; secondly, cold dark matter (CDM), which serves as the basis for the dark matter component. CDM is composed of non-relativistic, weakly interacting particles that are essentially separate from baryonic matter and radiation, primarily exerting their influence on the Universe through gravitational interactions. Despite the significant contributions of the $\Lambda$CDM model, standard cosmology faces several unresolved issues. The fine-tuning problem underscores the notable gap between the predicted theoretical values and the observed values of $\Lambda$ \cite{wein}. Furthermore, the coincidence problem raises an intriguing question: why do the energy densities of dark energy and dark matter, which evolve differently as the Universe expands, appear to have similar magnitudes in the present day? \cite{bis1}. These challenges indicate that a deeper understanding may be necessary to grasp the fundamental dynamics governing the Universe.\\
This intriguing scenario has been complicated by the emergence of what is termed the Hubble tension \cite{ht1}. This phrase describes a notable difference between the Hubble constant $H_0$ measured locally and the value derived from the Cosmic Microwave Background (CMB). Observations from the Hubble Space Telescope, which are calibrated using Cepheid variables, consistently provide a local measurement of $H_0 = 74 \pm 2 ~km s^{-1} Mpc^{-1}$ \cite{local}. Conversely, the CMB, through its measurements of temperature and polarization fluctuations that inform the free parameters in the $\Lambda$CDM model, suggests a lower value of $H_0 = 67.4 \pm 0.5~ km s^{-1} Mpc^{-1}$ \cite{early1}. To account for this inconsistency, various hypotheses regarding potential systematic errors in the measurements have been examined, but the data does not support these claims. This points toward the possibility of new physics that goes beyond the established cosmological framework \cite{sys}. As a result, this tension has motivated a variety of theoretical explorations, including changes to the physics governing both the early Universe and modifications to late-time dynamics. \\The latter approach is based on modification of expansion history of the Universe after the recombination period tend to focus on late-time phenomena. One method involves dynamic dark energy frameworks, which include models with a fluctuating equation of state. These models aim to align local and global measurements of $H_0$ but often exacerbate conflicts with other cosmological observations, such as growth rate data \cite{l1}. Another concept is based on chameleon fields, where local regions with high matter density capture a scalar field, thereby effectively boosting the local expansion rate \cite{l2}. However, numerous late-time modifications encounter obstacles, as they struggle to resolve the sound horizon issue or create discrepancies with baryon acoustic oscillation (BAO) and supernova observations \cite{l3}. A recent investigation also points out that late-time solutions face significant restrictions from inverse distance ladder techniques, which further diminish their feasibility \cite{l4}.\\
Solutions that focus on the early Universe generally involve alterations to the physics that existed prior to recombination. One captivating approach suggests the introduction of an unconventional energy component in the early Universe, known as Early Dark Energy (EDE) \cite{ede}. These models have shown potential in alleviating the tension by boosting the estimated $H_0$. This concept posits that EDE behaves similarly to a cosmological constant before the time of matter-radiation equality. In this approach, it is proposed that energy density of the Universe is elevated by approximately $10\%$ before recombination, leading to a reduction in the sound horizon at that time, which subsequently results in an increase in $H_0$ \cite{kom}. EDE is specifically designed to have an energy density that decreases more quickly than radiation during later periods, thereby ensuring that it does not influence the Universe's development after recombination.\\
In the present work, we will focus on the latter approach and consider a minimally coupled scalar field as EDE which directly couples with the radiation sector in the radiation-dominated era. This interacting scalar-photon model differs from recent interacting models \cite{tale} \cite{gom}, by establishing a direct link between the scalar field and radiation avoiding the need for symmetry breaking or disformal coupling that results in anisotropic expansion. Our analysis preserves isotropy and successfully establishes an effective cosmological constant during the radiation-dominated era without fine-tuning of the scalar field potential. This work is organized as follows: In Section 2, we outline the framework and derive the equations governing the fields within a cosmological model that assumes a flat Friedmann-Robertson-Walker cosmology. By establishing the equations related to the (non)-conservation of the scalar field and its interactions with radiation, we address the equation for energy density of radiation. Our findings indicate that the behavior of radiation energy density evolves differently from the conventional $\rho_{\gamma}\propto a^{-4}$ relationship, influenced by how energy is exchanged between EDE and the radiation. Moving on to Section 3, we illustrate that the model possesses two significant features associated with EDE scenarios. Firstly, it functions similarly to an effective cosmological constant, a characteristic that remains unchanged regardless of the potential. Secondly, the energy density attributed to the EDE component naturally decreases as the Universe continues to expand. In section 4, we outline our conclusions.

~~~~~~~~~~~~~~~~~~~~~~~~~~~~~~~~~~~~~~~~~~~~~~~~~~~~~~~~~~~~~~~~~~~~~~~~~~~~~~~~~~~~~~~~~~~~~~~~~~~~~~~~~~~~~~~~~~~~~~~~~~~~~~~~
\section{The model and field equations}
We consider the action functional
\begin{equation}
S= \int d^{4}x \sqrt{-g} \{\frac{1}{2}R
-\frac{1}{2}g^{\mu\nu}\nabla_{\mu}\phi
\nabla_{\nu}\phi-V(\phi)+C(\phi)L_{m}\}
\label{a1}\end{equation}
where $R$ is the Ricci scalar and $g$ is the determinant of the metric $g_{\mu\nu}$.
This action shows that a scalar field $\phi$ with the potential $V(\phi)$ interacts with Lagrangian density of matter $L_m$ via the coupling function $C(\phi)$. Such a setup arises in various extended gravitational theories. For instance, it is directly related to the Brans-Dicke (BD) theory where a scalar field emerges as a dynamical degree of freedom mediating the gravitational interaction.  In particular, within the BD theory, an exponential coupling function emerges naturally by going to Einstein frame via a conformal transformation. Similarly, Einstein frame  representations of $f(R)$ gravity yields a comparable structure where the scalar field effectively represents the modifications of General Relativity introduced by the $f(R)$ function. In fact, metric $f(R)$ gravity is dynamically equivalent to BD theory with a potential and a null BD parameter \cite{fr}. Moreover, due to a direct coupling of the scalar field with the matter sector the action (\ref{a1}) is also related to the so-called chameleon mechanism in which $\phi$ acquires an effective density dependent mass \cite{cham1}.\\
By taking the action (\ref{a1}) as the starting point, we would like to consider a cosmological model involving a scalar field coupled to radiation at early times before recombination. Our main goal is to study how such an interacting system affects the expansion rate of the Universe at early times\footnote{Such an interacting scalar-photon system has been also used to study its effect on the redshift evolution of the CMB
temperature and varying fundamental constants\cite{avgo}.}. In particular, we will focus on answering the question whether the interaction can increase the energy density of the radiation so that the sound horizon at recombination is decreased and potentially alleviate the Hubble tension.\\
To do this, we consider the action (\ref{a1}) with an exponential coupling function $C(\phi)=e^{-\sigma\phi}$ with $\sigma$ being a coupling parameter\footnote{In Einstein frame representation of BD theory, this parameter is related to the BD parameter \cite{bis2}.}. For a spatially flat FRW metric, the conservation equations then become\footnote{We have chosen $L_m=p_m$ for the Lagrangian density \cite{bis2}.},
\begin{equation}
\dot{\rho}_{\phi}+3H(\omega_{\phi}+1)\rho_{\phi}=-\frac{1}{3}\sigma e^{-\sigma\phi}\dot{\phi}\rho_{\gamma}
\label{a2}\end{equation}
\begin{equation}
\dot{\rho}_{\gamma}+4H\rho_{\gamma}=\frac{4}{3}\sigma \dot{\phi}
\rho_{\gamma} \label{a3}\end{equation}
where $\omega_{\phi}=\frac{\rho_{\phi}}{p_{\phi}}$, $\rho_{\phi}=\frac{1}{2}\dot{\phi}+V(\phi)$, $p_{\phi}=\frac{1}{2}\dot{\phi}-V(\phi)$ and
$\rho_{\gamma}$ is the energy density of radiation. The Hubble parameter $H$ satisfies
\begin{equation}
3H^2= \rho_{eff}= e^{-\sigma\phi}\rho_{\gamma}+\rho_{\phi}
\label{a4}\end{equation}
\begin{equation}
2\dot{H}+3H^2= -p_{eff}=-e^{-\sigma\phi}p_{\gamma}-p_{\phi}
\label{a5}\end{equation}
with $p_{\gamma}=\frac{1}{2}\rho_{\gamma}$. The equation (\ref{a3}) can immediately be solved which gives \cite{bis2}
\begin{equation}
\rho_{\gamma}=\rho_{0\gamma}a^{-4+\epsilon}
\label{a6}\end{equation} with
\begin{equation}
\epsilon\equiv \frac{4\sigma\phi}{3\ln a}\label{a16}\end{equation}
and $\rho_{0\gamma}$ being an integration constant. The energy transfer between $\phi$ and radiation
is parameterized by $\epsilon$. For $\epsilon>0$, radiation is created and energy is injecting from $\phi$ into radiation so that the latter dilutes more slowly compared to
the standard evolution $\rho_{\gamma}\propto a^{-4}$. Conversely, when $\epsilon<0$ radiation is annihilated
and energy transfers outside of radiation so that the rate of dilution of $\rho_{\gamma}$ is faster than the
standard one. Taking $\epsilon$ as a constant parameter then $\phi$ evolves logarithmically with the scale factor
\begin{equation}
\phi=\gamma\ln a
\label{a7}\end{equation}
with $\gamma\equiv \frac{3\epsilon}{4\sigma}$. The relation (\ref{a7}) implies that the rate of change of $\phi$ is given by the Hubble parameter, namely that $\dot{\phi}=\gamma H$.
~~~~~~~~~~~~~~~~~~~~~~~~~~~~~~~~~~~~~~~~~~~~~~~~~~~~~~~~~~~~~~~~~~~~~~~~~~~~~~~~~~~~~~~~~~~~~~~~~~~~~~~~~~~~~~~~~~~~~~~~~~~~
\section{The scalar field $\phi$ as EDE}
There are two important requirements that play an essential role in EDE scenarios \cite{1}. First, the EDE component should have an equation of state parameter that behaves similarly to that of a cosmological constant at early times, playing a significant role in the speeding up of expansion of the Universe. Second, it is essential that the energy density of the EDE component decreases more rapidly than that of radiation as the Universe grows. This rapid decrease guarantees that, before recombination occurs, the EDE component becomes less significant which in turn helps to maintain the typical processes of structure formation in subsequent stages. These dynamics are vital for aligning with observations of CMB and the large-scale structure of the Universe.\\
Let us check the first requirement for the above model. From the equations (\ref{a4}) and (\ref{a5}), we can write for the effective fluid
\begin{equation}
\omega_{eff}\equiv \frac{p_{eff}}{\rho_{eff}}
=\frac{ e^{-\sigma\phi}p_{\gamma}+p_{\phi}}{e^{-\sigma\phi}\rho_{\gamma}+\rho_{\phi}}
\label{a11}\end{equation}
Combining the latter with (\ref{a4}) and (\ref{a7}) gives
\begin{equation}
\omega_{eff}=(1-\frac{1}{3}\gamma^2)\frac{\frac{1}{3}e^{-\sigma\phi}\rho_{\gamma}-V(\phi)}{e^{-\sigma\phi}\rho_{\gamma}+V(\phi)}
\label{a12}\end{equation}
At early times when $a\rightarrow 0$, $\phi$ goes to infinity as inferred by the relation (\ref{a7}). This means that  $e^{-\sigma\varphi}\rho_{r}\rightarrow 0$ in (\ref{a12}) and then $\omega_{eff}$ takes a constant value $\omega_{eff}\rightarrow \frac{1}{3}\gamma^2-1$ which is bounded by $-1<\omega_{eff}<0$ for $\gamma<1$. In particular, for $\gamma<<1$ (or $\epsilon<<\frac{4}{3}\sigma$) we have $\omega_{eff}\approx -1$ meaning that the effective fluid behaves like a cosmological constant. It should be emphasized that this behavior is independent of the shape of the potential $V(\phi)$ and the scalar field equation of state parameter.\\
For checking the second requirement, we solve the equation (\ref{a2}) under the assumption that $\omega_{\phi}\equiv\alpha\approx const$. This assumption simplifies the equation allowing us to express
$\rho_{\phi}$ as a function of the scale factor. In this case, we can rewrite (\ref{a2}) in the form
\begin{equation}
\dot{r}+[(\epsilon-4)+3(\alpha+1)] Hr=-\frac{\epsilon}{4} H a^{-\frac{3\epsilon}{4}}
\label{a8}\end{equation}
where (\ref{a7}) is used and $r\equiv\frac{\rho_{\phi}}{\rho_{\gamma}}$  measures the fractional energy densities of $\phi$
and radiation. The equation (\ref{a8}) can be rewritten as
\begin{equation}
\frac{dr}{da}+[(\epsilon-4)+3(\alpha+1)]\frac{r}{a}=-\frac{\epsilon}{4} a^{-\frac{3\epsilon}{4}-1}
\label{a8-1}\end{equation}
The solution is
\begin{equation}
r(a)=\frac{\epsilon}{4\lambda}a^{-\frac{3\epsilon}{4}}
+C a^{-\lambda-\frac{3\epsilon}{4}}
\label{a9}\end{equation}
with $\lambda\equiv(3\alpha-1)+\frac{\epsilon}{4}$ and $C$ being an integration constant. We would like to determine the constant $C$ by imposing a boundary condition at some critical value for the scale factor at early times (denoted by $a_c$) which is presumed to occur shortly after the inflationary period. We impose the condition that at this early epoch, $r(a\rightarrow a_c)\rightarrow 1$. This condition reflects a cosmological scenario where the scalar field and radiation contribute equally to the total energy density at the scale factor $a_c$. With this boundary condition, (\ref{a9}) becomes
\begin{equation}
r(a)=\frac{\epsilon}{4\lambda}a^{-\frac{3\epsilon}{4}}\{1+[\frac{4\lambda}{\epsilon}a^{\frac{3\epsilon}{4}}-(\frac{a}{a_c})^{\frac{3\epsilon}{4}}
](\frac{a}{a_c})^{-\lambda-{\frac{3\epsilon}{4}}}\}
\label{a10}\end{equation}
When the scale factor becomes sufficiently larger than $a_c$ ($a>>a_c$), $(\frac{a}{a_c})^{-\lambda-\frac{3\epsilon}{4}}\approx 0$ for positive $\lambda$ and $\epsilon$. Consequently, the solution (\ref{a10}) simplifies to $r(a)\approx \frac{\epsilon}{4\lambda}a^{-\frac{3\epsilon}{4}}$. This suggests that when $\epsilon>0$\footnote{Note that this corresponds to the case that energy transfers from $\phi$ to radiation. }, the function $r(a)$ tends to decline as the Universe continues to expand. In physical terms, this situation reflects a shift of energy from the scalar field $\phi$ to the radiation segment of the Universe. As a result, the contribution of $\rho_{\phi}$ becomes less significant compared to $\rho_{\gamma}$ over time. As stated above, this process is crucial in cosmological theories where it is anticipated that the energy density of the scalar field will decrease faster than that of radiation. This also ensures that radiation remains the dominant energy component in the Universe before the period of recombination.
\begin{figure}[h]
\begin{center}
\includegraphics[width=0.4\linewidth]{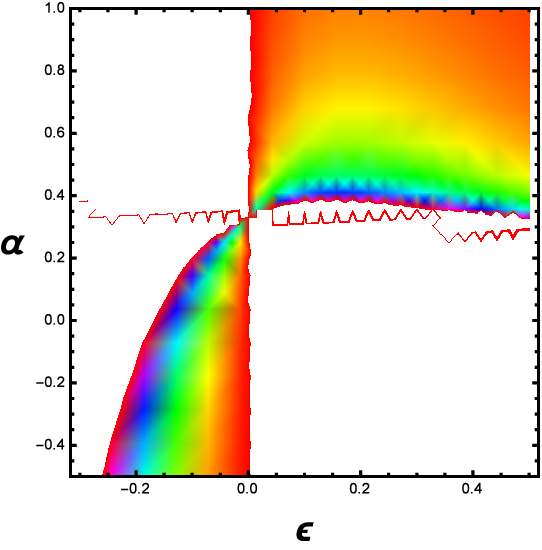}
\includegraphics[width=0.06\linewidth]{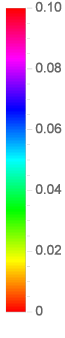}
\caption{Density plot of $r(a)$ in the parameters space ($\alpha$, $\epsilon$). This plot highlights the regions in which the ratio
$r(a)$ lies between $0$ and $0.1$, corresponding to a scenario where EDE contributes up to $10\%$ of the total energy budget of the Universe before recombination.}
\end{center}
\end{figure}
The fig.1 shows a density plot of $r(a)$ in the parameters space ($\alpha$, $\epsilon$) close to the recombination period ($a \approx 1100$). It illustrates the regions in the parameters space where $r(a)$ accounts for up to $10\%$ of the Universe's total energy content. This aligns with recent findings that limit the EDE contribution to the overall energy density at the time of recombination \cite{kom}. Consequently, the figure offers a visual depiction of the permissible parameters range for EDE models, ensuring their alignment with existing cosmological observations.
~~~~~~~~~~~~~~~~~~~~~~~~~~~~~~~~~~~~~~~~~~~~~~~~~~~~~~~~~~~~~~~~~~~~~~~~~~~~~~~~~~~~~~~~~~~~~~~~~~~~~~~~~~~~~~~~
\section{Conclusions}
In this study, we have constructed a model of EDE in which a minimally coupled scalar field, representing the EDE, interacts with the radiation sector through an exponential coupling function given by $C(\phi) = e^{-\sigma \phi}$. Our examination indicates that this interaction alters the traditional law of radiation conservation, resulting in a modified scaling relation $\rho_{\gamma} \propto a^{-4 + \epsilon}$, where the parameter $\epsilon$ measures the energy transfer between radiation and EDE. Even though this quantity may generally evolves with expansion, we considered it as a constant parameter. This simplifies our analysis and gives the scalar field evolution according to $\phi \propto \ln a$ suggesting that $\phi$ diverges as $a$ approaches zero at early times. In our analysis, $\epsilon>0$ corresponds to transferring energy from EDE to radiation
which causes energy density of the latter increases before recombination so that the sound horizon at recombination is decreased leading to alleviation of the Hubble tension.\\
We have shown that the effective fluid, made up of EDE and radiation, exhibits an equation of state of an effective cosmological constant during early stages of evolution of the Universe.
Additionally, we analytically solved the conservation equation for the scalar field, assuming a constant equation-of-state parameter $\omega_{\phi}\equiv \alpha$ which led us to derive an expression for the ratio $r(a) = \frac{\rho_{\phi}}{\rho_{\gamma}}$. Our findings show that for feasible parameter selections, especially with $\epsilon > 0$ (indicating energy transfer from EDE to radiation), this ratio diminishes as the Universe expands. This characteristic is crucial for any EDE model, as it guarantees that while the EDE component momentarily boosts the expansion of the Universe (thereby compressing the sound horizon and potentially addressing the Hubble tension), its effect becomes insignificant before recombination, avoiding disruptions to the well-established processes of structure formation in subsequent epochs.\\
Furthermore, we investigated the parameters space defined by ($\alpha$, $\epsilon$) and showed that the anticipated behavior of $r(a)$ around the time of recombination aligns with contemporary observational data, thus meeting one of the essential criteria for a successful EDE model. In summary, our research offers a solid framework that not only broadens the scaling solutions identified in non-interacting models but also provides fresh perspectives on how interactions within the dark sector can harmonize the dynamics of the early Universe with observations from later times. Future investigations will aim to refine these parameter constraints through in-depth numerical analyses and to compare our predictions with forthcoming CMB and large-scale structure observations.
~~~~~~~~~~~~~~~~~~~~~~~~~~~~~~~~~~~~~~~~~~~~~~~~~~~~~~~~~~~~~~~~~~~~~~~~~~~~~~~~~~~~~~~~~~~~~~~~~~~~~~~~~~~~~~~~

\end{document}